\documentstyle[prl,aps,preprint]{revtex}    
\begin{document}
\draft

\tightenlines 
\renewcommand{\theequation}{\arabic{section}$\cdot$\arabic{equation}}
\def\diff{{\rm d}}
\def\bfk{{\bf k}}
\def\eps{\epsilon}
\def\bra#1{{\langle #1 |}}     
\def\ket#1{{| #1 \rangle}}     
\def\maru#1{{\check #1}}
\def\mib#1{\mbox{\boldmath $#1$}}
\title{
ON A POSSIBLE SCHEME OF q-DEFORMATION OF 
MANY-BOSON SYSTEMS IN 
TIME-DEPENDENT VARIATIONAL METHOD
\footnote{Talk presented by M.Y. at XXVI International Workshop on 
Condensed Matter Theories, Luso, Portugal, 2-7 September 2002. 
To appear in CMT26 (Luso) Workshop Proceedings Vol.18, 
Condensed Matter Theories, Nova Science Publishers.}
}
\author{
Atsushi Kuriyama,$^1$ Constan\c{c}a Provid\^encia,$^2$ 
Jo\~ao da Provid\^encia,$^2$ Yasuhiko Tsue$^3$ and Masatoshi Yamamura$^1$
}
\address{
$^1$ Faculty of Engineering, Kansai University, 
Suita 564-8680, Japan} 
\address{
$^2$ Departamento de Fisica, Universidade de Coimbra, 
P-3000 Coimbra, Portugal}
\address{
$^3$ Physics Division, Faculty of Science, 
Kochi University, Kochi 780-8520, Japan} 
\date{\today} 

\maketitle

\begin{abstract}
A possible scheme of $q$-deformation, which was recently developed by the 
present authors, is reviewed by stressing the starting idea. 
\end{abstract}

\narrowtext

\newpage

\section{INTRODUCTION}

Time-dependent variational method may be one of the powerful methods for 
investigating time-evolution of quantum mechanical systems in the framework 
of a possible approximation. In this method, a trial state for the variation 
is prepared as a function of variational parameters. 
If a certain condition is introduced, the time-dependent variational method 
can be formulated in the framework of classical Hamiltonian mechanics 
\cite{1}. 

For many-boson systems consisting of a kind of boson operator, the simplest 
trial state may be boson coherent state. Ref.\cite{1} tells us that the boson 
coherent state is an eigenstate of boson annihilation operator with 
the eigenvalue which is regarded as a canonical variable in classical 
mechanics. Further, we know that the boson coherent state consists of the 
states with the boson numbers from 0 to $\infty$ in a fixed superposition. 
In order to give variety to the trial state, the boson coherent state 
must be generalized. This generalized boson coherent state consists of 
the states of the boson numbers from 0 to $\infty$ with a superposition 
different from the boson coherent state. As for the idea for the 
generalization, there exists boson annihilation like operator, the 
eigenstate of which is just the generalized boson coherent state and the 
eigenvalue is the variational parameter. The above idea can be realized 
through introducing a function of the boson number operator. 
This boson annihilation like operator and its hermit conjugate 
form $q$-deformation of the boson system. By adopting various function 
of the boson number operator, we are able to obtain various types of the 
$q$-deformation. The concept of the $q$-deformation serves us an interesting 
viewpoint for understanding dynamics of many-body systems described by 
boson operators. In this sense, it may be interesting to investigate the 
$q$-deformation of many-boson systems in relation to the generalized boson 
coherent states as the trial states in the time-dependent variational 
method.

The aim of this report is to present the starting idea for the 
$q$-deformation based on the above-mentioned scheme which was recently 
developed by the present authors \cite{2}. 
Of course, the background of the present 
report is colored with the papers mainly published by the 
Coimbra group \cite{3}. 
Of course, the paper by Penson and Solomon \cite{4} also gave us independent 
influence on the present work.

In \S\S 2 and 3, the time-dependent variational method is recapitulated. 
Section 4 is a central part, in which three forms of generalized boson 
coherent states and $q$-deformation are discussed. In \S 5, the 
Holstein-Primakoff boson representation for the $su(2)$- and 
$su(1,1)$-algebras is shown as a result of the $q$-deformation. In \S 6, 
as a simple example, damped and amplified oscillation is discussed 
in the language of the $q$-deformation. Finally, in \S 7, two remarks 
are mentioned.

\setcounter{equation}{0}

\section{TIME-DEPENDENT VARIATIONAL METHOD FORMULATED 
IN TERMS OF CLASSICAL CANONICAL VARIABLES}

First, let us recapitulate the time-dependent variational method 
formulated in Ref.\cite{1}. 
For simplicity, a trial state for the time-dependent variation is 
parametrized in terms of two real parameters ($p, q$). The trial state is 
denoted by $\ket{p,q}$. 
Let $\ket{p,q}$ obey the following condition : 
\begin{equation}\label{2-1}
\bra{p,q}i\partial_{q}\ket{p,q}=p+\partial_q S(p,q) \ , 
\qquad
\bra{p,q}i\partial_{p}\ket{p,q}=\partial_p S(p,q) \ . 
\end{equation}
Next, the following quantity is introduced : 
\begin{equation}\label{2-2}
L=\bra{p,q}i\partial_t -{\hat H}\ket{p,q} \ . 
\end{equation}
Here, ${\hat H}$ denotes the Hamiltonian under investigation. With the 
use of the condition (\ref{2-1}), $L$ can be rewritten as 
\begin{eqnarray}
& &
L=p{\dot q}-H(p,q)+{\dot S}(p,q) \ , 
\label{2-3}\\
& &
H(p,q)=\bra{p,q}{\hat H}\ket{p,q}\ . 
\label{2-4}
\end{eqnarray}

The time-dependent variational method is formulated in the following form :
\begin{equation}\label{2-5}
\delta\int L dt =0 \ .
\end{equation}
The relation (\ref{2-5}), together with the form (\ref{2-3}), 
gives us the following equation : 
\begin{equation}\label{2-6}
{\dot q}=\partial_p H  \ , \qquad
{\dot p}=-\partial_q H \ .
\end{equation}
The relation (\ref{2-6}) is nothing but the Hamilton's equation of motion. 
Of course, $L$ and $H$ denote the Lagrangian and the Hamiltonian, 
respectively, and $(p, q)$ can be regarded as the canonical variable in 
classical mechanics. 
By solving Eq.(\ref{2-6}) under an appropriate initial condition, 
the time-dependence of $(p, q)$ is determined. 
Then, $\ket{p,q}$ is obtained as a function of $t$ and 
the time-evolution of the system under investigation 
is described in the framework of the time-dependent variational method.

\setcounter{equation}{0}

\section{BOSON COHERENT STATE AS A TRIAL STATE 
FOR THE VARIATION}

Instead of the variable $(p,q)$, the following ones denoted as 
$(z, z^*)$ is useful in some cases : 
\begin{equation}\label{3-1}
p=i(z^*-z)/\sqrt{2} \ , \qquad
q=(z^*+z)/\sqrt{2} \ .
\end{equation}
Then, for the case $S=pq/2$, the condition (\ref{2-1}) can be rewritten as 
\begin{eqnarray}
& &
\bra{p,q}\partial_z\ket{p,q}=+z^*/2 \ , 
\qquad
\bra{p,q}\partial_{z^*}\ket{p,q}=-z/2 \ , 
\label{3-2}\\
& &
S=i(z^{*2}-z^2)/4 \ . 
\label{3-3}
\end{eqnarray}
The Lagrangian $L$ can be expressed in the form 
\begin{equation}\label{3-4}
L=i(z^*{\dot z}-z{\dot z}^*)-H(z,z^*) \ .
\end{equation}
The time-dependent variation gives us 
\begin{equation}\label{3-5}
i{\dot z}=+\partial_{z^*}H \ , \qquad
i{\dot z}^*=-\partial_z H \ .
\end{equation}

As a possible trial state, let us adopt the boson coherent state 
in the form 
\begin{equation}\label{3-6}
\ket{p,q}=\ket{c}=\left(\sqrt{\Gamma}\right)^{-1}
\exp\left(\gamma{\hat c}^*\right)\ket{0} \ .
\end{equation}
Here, $\Gamma$ denotes the normalization constant : 
\begin{equation}\label{3-7}
\Gamma=\exp\left(|\gamma|^2\right) \ .
\end{equation}
The state $\ket{c}$ satisfies 
\begin{equation}\label{3-8}
{\hat \gamma}={\hat c} \ , \qquad
{\hat \gamma}\ket{c}=\gamma\ket{c} \ .
\end{equation}
The condition (\ref{3-2}) can be rewritten as 
\begin{eqnarray}\label{3-9}
& &
\bra{c}\partial_z \ket{c}=(\gamma^*\cdot \partial_z \gamma-
\gamma\cdot \partial_z \gamma^*)/2 = +z^*/2 \ , 
\nonumber\\
& &
\bra{c}\partial_{z^*}\ket{c}=(\gamma^*\cdot\partial_{z^*}\gamma
-\gamma\cdot\partial_{z^*}\gamma^*)/2=-z/2 \ .
\end{eqnarray}
The condition (\ref{3-9}) gives us 
\begin{equation}\label{3-10}
\gamma=z \ , \qquad \gamma^*=z^* \ .
\end{equation}
The parameter $(\gamma, \gamma^*)$ can be regarded as boson type canonical 
variable.

\setcounter{equation}{0}

\section{THREE FORMS OF GENERALIZED BOSON COHERENT STATES 
AND $q$-DEFORMATION}

In order to generalize the trial states from the boson coherent state, 
let us introduce the following three states : 
\begin{eqnarray}
& &\ket{c_0}=\left(\sqrt{\Gamma_0}\right)^{-1}
\exp\left(\gamma_0{\hat c}^* f_0({\hat N})\right)\ket{0} \ , 
\label{4-1a}\\
& &\ket{c_P}=\left(\sqrt{\Gamma_P}\right)^{-1}{\hat P}_{n^0}\cdot
\exp\left(\gamma_P{\hat c}^* f_P({\hat N})\right)\ket{0} \ , 
\label{4-1b}\\
& &\ket{c_Q}=\left(\sqrt{\Gamma_Q}\right)^{-1}{\hat Q}_{n^0}\cdot
\exp\left(\gamma_Q{\hat c}^* f_Q({\hat N})\right)\ket{0} \ , 
\label{4-1c}\\
& &{\hat N}={\hat c}^*{\hat c} \ , 
\label{4-2}\\ 
& &{\hat P}_{n^0}=\sum_{n=0}^{n^0} \ket{n}\bra{n} \ , \qquad
{\hat Q}_{n^0}=1-P_{n^0} \ , 
\label{4-3}\\
& &\ket{n}=\left(\sqrt{n!}\right)^{-1}({\hat c}^*)^n\ket{0} \ . 
\label{4-4}
\end{eqnarray}
The function $f_R({\hat N})$ $(R=0, P, Q)$ is defined by the relation 
\begin{equation}\label{4-5}
f_R({\hat N})\ket{n}=f_R(n)\ket{n} \ . \quad (R=0, P, Q)
\end{equation}
For the function $f_R(n)$, the following conditions are imposed:
\begin{eqnarray}
& &f_0(0)=1 \ , \qquad f_0(n)>0\quad (n=1,2,3,\cdots) \ , 
\label{4-6a}\\
& &f_P(0)=1 \ , \qquad f_P(n)>0 \quad (n=1,2,\cdots, n^0-1) \ , 
\nonumber\\
& &f_P(n) \ ; \ \hbox{\rm arbitrary} \ (n=n^0, n^0+1,\cdots) \ , 
\label{4-6b}\\
& &f_Q(n)=1 \quad (n=0,1,\cdots,n^0) \ , 
\nonumber\\
& &f_Q(n)>0 \quad (n=n^0+1, n^0+2,\cdots) \ . 
\label{4-6c}
\end{eqnarray}

Next, the following operators are introduced : 
\begin{equation}
{\hat \gamma}_0=f_0({\hat N})^{-1}{\hat c} \ , 
\qquad
{\hat \gamma}_P={\hat P}_{n^0}f_P({\hat N})^{-1}{\hat c} \ , 
\qquad
{\hat \gamma}_Q={\hat Q}_{n^0}f_Q({\hat N})^{-1}{\hat c} \ .
\label{4-7}
\end{equation}
They obey the relations 
\begin{equation}
{\hat \gamma}_0\ket{c_0}=\gamma_0\ket{c_0} \ , 
\qquad
{\hat \gamma}_P\ket{c_P}=\gamma_P{\hat P}_{n^0}\ket{c_P} \ ,
\qquad
{\hat \gamma}_Q\ket{c_Q}=\gamma_Q\ket{c_Q} \ . \ \ \ 
\label{4-8}
\end{equation}
The forms 
(\ref{4-8}) 
permit us to call the states 
$\ket{c_0}$, $\ket{c_P}$ and $\ket{c_Q}$ the generalized coherent states. 
The operator $({\hat \gamma}_0 , {\hat \gamma}_0^*)$ can be 
regarded as $q$-deformed boson operator which is characterized by the 
function $f_0({\hat N})$. It is justified through the following relations : 
\begin{eqnarray}
& &\ket{n}=\left(\sqrt{[n]_q!}\right)^{-1}({\hat \gamma}_0^*)^n\ket{0} \ .
\quad (n=0,1,2,\cdots) \nonumber\\
& & {\hat \gamma}_0\ket{0}=0 \ , 
\label{4-9}\\
& &{\hat \gamma}_0^*{\hat \gamma}_0=[{\hat N}]_q \ , 
\qquad {\hat \gamma}_0{\hat \gamma}_0^*=[{\hat N}+1]_q \ , 
\label{4-10}\\
& &{\hat \gamma}_0\ket{n}=\sqrt{[n]_q}\ket{n-1} \ , 
\quad\ \ 
{\hat \gamma}_0^*\ket{n}=\sqrt{[n+1]_q}\ket{n+1} \ , 
\label{4-11}\\
& &[{\hat N} , {\hat \gamma}_0]=-{\hat \gamma}_0 \ , \quad
[{\hat N} , {\hat \gamma}_0^*]=+{\hat \gamma}_0^* \ , 
\quad
{\hat N}\ket{n}=n\ket{n}\ . 
\label{4-12}
\end{eqnarray}
The quantities $[n]_q!$ and $[x]_q$ are defined as 
\begin{eqnarray}
& &[n]_q!=n!(f_0(0)\cdots f_0(n-1))^{-2} \ , \qquad\qquad \qquad\qquad\quad
[0]_q!=1 \ , 
\label{4-13}\\
& &[x]_q=xf_0(x-1)^{-2} \ , \quad (x={\hat N}, {\hat N}+1, n, n+1) \ , \qquad
[0]_q=0 \ .
\label{4-14}
\end{eqnarray}

From the relation (\ref{4-10}), the following relation is derived : 
\begin{eqnarray}
{\hat \gamma}_0{\hat \gamma}_0^*-(f_0({\hat N}-1)^2f_0({\hat N})^{-2}
+F({\hat N})){\hat \gamma}_0^*{\hat \gamma}_0 
&=&f_0({\hat N})^{-2}-F({\hat N}){\hat N}f_0({\hat N}-1)^{-2} \ , 
\nonumber\\
& &\qquad F({\hat N}) \ : \ {\rm arbitrary} \ . 
\label{4-15}
\end{eqnarray}
Especially, the commutation relation for ${\hat \gamma}_0$ and 
${\hat \gamma}_0^*$ is given in the form 
\begin{equation}\label{4-16}
[{\hat \gamma}_0 , {\hat \gamma}_0^*]=[{\hat N}+1]_q-[{\hat N}]_q
=\Delta_N({\hat \gamma}_0^*{\hat \gamma}_0) \ .
\end{equation}
Here, $\Delta_N({\hat \gamma}_0^*{\hat \gamma}_0)$ denotes the 
difference with $\Delta N=1$. 

We show three concrete examples :  

\noindent
(1) The most popular form : 
\begin{eqnarray}
& &f_0(n)=\sqrt{(n+1)(q-q^{-1})/(q^{n+1}-q^{-(n+1)})} \ , 
\label{4-17}\\
& &{\hat \gamma}_0{\hat \gamma}_0^*-q^{-1}{\hat \gamma}_0^*{\hat \gamma}_0
=q^{{\hat N}} \ , \nonumber\\
& &F({\hat N})=q^{-1}-f_0({\hat N}-1)^{2}f_0({\hat N})^{-2} \ .
\label{4-18}
\end{eqnarray}

\noindent
(2) The form presented by Penson and Solomon \cite{4} : 
\begin{eqnarray}
& &f_0(n)=q^{n/2} \ , 
\label{4-19}\\
& &{\hat \gamma}_0{\hat \gamma}_0^*-q^{-1}{\hat \gamma}_0^*{\hat \gamma}_0
=q^{-{\hat N}} \ , 
\nonumber\\
& &F({\hat N})=0 \ . 
\label{4-20}
\end{eqnarray}

\noindent
(3) A possible modification given by the present authors : 
\begin{eqnarray}
& &f_0(n)=\sqrt{(n+1)(1-q^{-2})/(1-q^{-2(n+1)})} \ , 
\label{4-21}\\
& &{\hat \gamma}_0{\hat \gamma}_0^*-q^{-2}{\hat \gamma}_0^*{\hat \gamma}_0 
=1 \ , 
\nonumber\\
& &F({\hat N})=q^{-1}-f_0({\hat N}-1)^2f_0({\hat N})^{-2} \ . 
\label{4-22}
\end{eqnarray}

We can formulate classical counterpart of the $q$-deformation. 
For the state $\ket{c_0}$, the following relation can be derived : 
\begin{eqnarray}
& &\bra{c_0}\partial_z\ket{c_0}
=(\gamma_0^*\partial_z\gamma_0-\gamma_0\partial_z\gamma_0^*)/2\cdot
(\Gamma_0'/\Gamma_0)=+z^*/2 \ , 
\nonumber\\
& &\bra{c_0}\partial_{z^*}\ket{c_0}
=(\gamma_0^*\partial_{z^*}\gamma_0-\gamma_0\partial_{z^*}\gamma_0^*)/2
\cdot(\Gamma_0'/\Gamma_0)=-z/2 \ , 
\label{4-23}\\
& &\qquad\qquad\qquad
\Gamma_0'=d\Gamma_0/d|\gamma_0|^2 \ .
\label{4-24}
\end{eqnarray}
The condition (\ref{4-23}) gives us 
\begin{equation}\label{4-25}
z=\gamma_0\sqrt{\Gamma_0'/\Gamma_0} \ , \qquad
z^*=\gamma_0^*\sqrt{\Gamma_0'/\Gamma_0} \ .
\end{equation}
Since $\Gamma_0$ is a function of $|\gamma_0|^2$, $(\gamma_0 , \gamma_0^*)$ 
can be expressed in terms of $(z, z^*)$. 
Then, the Hamilton's equation of motion is derived : 
\begin{equation}\label{4-26}
i{\dot z}=\partial_{z^*}H \ , \quad
i{\dot z}^*=-\partial_z H \ , \qquad
H=\bra{c_0}{\hat H}\ket{c_0} \ .
\end{equation}
However, actually, in many cases, it is impossible to express 
$(\gamma_0 , \gamma_0^*)$ in terms of $(z, z^*)$.

The state $\ket{c_0}$ satisfies the relations 
\begin{eqnarray}
& &
\bra{c_0}{\hat \gamma}_0\ket{c_0}=\gamma_0 \ , \qquad
\bra{c_0}{\hat \gamma}_0^*\ket{c_0}=\gamma_0^* \ , 
\label{4-27}\\
& &\bra{c_0}{\hat \gamma}_0^*{\hat \gamma}_0\ket{c_0}=\gamma_0^*\gamma_0 \ .
\label{4-28}
\end{eqnarray}
Further, $\bra{c_0}{\hat N}\ket{c_0}$ is given as 
\begin{equation}\label{4-29}
\bra{c_0}{\hat N}\ket{c_0}=\gamma_0^*\gamma_0\cdot \Gamma_0'/\Gamma_0
=z^*z\ (=N) \ .
\end{equation}
With the use of the relation (\ref{4-26}), the following relations 
are derived : 
\begin{eqnarray}
& &[N , \gamma_0]_P=-\gamma_0 \ , \qquad
[N , \gamma_0^*]_P=\gamma_0^* \ , 
\label{4-30}\\
& &[\gamma_0 , \gamma_0^* ]_P=d_N(\gamma_0^*\gamma_0) \ .
\label{4-31}
\end{eqnarray}
Here, $d_N(\gamma_0^*\gamma_0)$ denotes the differential with respect to 
$N$ and $[A, B]_P$ expresses the Poisson bracket : 
\begin{equation}\label{4-32}
[ A , B]_P=\partial_zA\cdot\partial_{z^*}B-\partial_{z^*}A\cdot
\partial_zB \ .
\end{equation}
%

Thus, the following correspondence is obtained : 
\begin{equation}\label{4-33}
{\hat \gamma}_0 \sim \gamma_0 \ , \qquad
{\hat \gamma}_0^*\sim\gamma_0^* \ , \qquad
{\hat N}\sim N \ .
\end{equation}
The Hamilton's equation is written as 
\begin{eqnarray}
& &
i{\dot \gamma}_0=+\partial_{\gamma_0^*}H\cdot(N')^{-1} \ , 
\quad
i{\dot \gamma}_0^*=-\partial_{\gamma_0}H\cdot (N')^{-1}\ , 
\label{4-34}\\
& &\qquad
N'=dN/d|\gamma_0|^2 \ .
\label{4-35}
\end{eqnarray}

\setcounter{equation}{0}

\setcounter{equation}{0}

\section{THE HOLSTEIN-PRIMAKOFF BOSON REPRESENTATION FOR 
THE $su(2)$- AND $su(1,1)$-ALGEBRAS AS $q$-DEFORMATION}

Under appropriate choices of $f_R(n)$ ($R=0,P,Q$), the $q$-deformation 
leads us to the Holstein-Primakoff boson representation for the 
$su(2)$- and $su(1,1)$-algebras. We show three cases. 
$$
(1)\qquad
f_0(n)=\left(\sqrt{1+n/n^0}\right)^{-1} \ : \ (n^0 \ : \ {\rm positive}) 
\qquad\qquad\qquad\qquad\qquad\qquad
$$
\begin{eqnarray}\label{5-1}
& &{\hat \gamma}_0=(\sqrt{n^0})^{-1}\cdot {\hat T}_- \ , \qquad
{\hat T}_-=\sqrt{n^0+{\hat N}}\ {\hat c} \ , 
\nonumber\\
& &{\hat \gamma}_0^*=(\sqrt{n^0})^{-1}\cdot {\hat T}_+ \ , \qquad
{\hat T}_+={\hat c}^*\sqrt{n^0+{\hat N}} \ , 
\nonumber\\
& &[{\hat \gamma}_0 , {\hat \gamma}_0^*]=2(n^0)^{-1}\cdot{\hat T}_0 \ , 
\quad 
{\hat T}_0={\hat N}+n^0/2 \ .
\end{eqnarray}
The form of ${\hat T}_{\pm,0}$ is identical to the Holstein-Primakoff 
boson representation of the $su(1,1)$-algebra. 
%
$$
(2)\qquad
f_P(n)=\left(\sqrt{1-n/n^0}\right)^{-1} \ : \ (n^0 \ : \ {\rm positive}) 
\qquad\qquad\qquad\qquad\qquad\qquad
$$
\vspace{-1cm}
\begin{eqnarray}\label{5-2}
& &{\hat \gamma}_P=(\sqrt{n^0})^{-1}\!\cdot\!
{\hat P}_{n^0}\!\cdot\!{\hat S}_-\!\cdot\!{\hat P}_{n^0} \ , 
\qquad
{\hat S}_-=\sqrt{n^0-{\hat N}}\ {\hat c}\ , 
\nonumber\\
& &{\hat \gamma}_P^*=(\sqrt{n^0})^{-1}\!
\cdot\!{\hat P}_{n^0}\!\cdot\!{\hat S}_+\!\cdot\!{\hat P}_{n^0} \ , \qquad
{\hat S}_+={\hat c}^*\sqrt{n^0-{\hat N}} \ , 
\nonumber\\
& &[{\hat \gamma}_P^* , {\hat \gamma}_P ]=2(n^0)^{-1}\!\cdot\!
{\hat P}_{n^0}\!\cdot\!{\hat S}_0\!\cdot\!{\hat P}_{n^0} \ , \quad
{\hat S}_0={\hat N}-n^0/2 \ .
\end{eqnarray}
The form of ${\hat S}_{\pm, 0}$ is identical to the Holstein-Primakoff 
boson representation of the $su(2)$-algebra.
%
$$
(3)\qquad
f_Q(n)=\left(\sqrt{n/n^0-1}\right)^{-1} \ : \ (n^0 \ : \ {\rm positive}) 
\qquad\qquad\qquad\qquad\qquad\qquad
$$
\vspace{-1cm}
\begin{eqnarray}\label{5-3}
& &{\hat \gamma}_Q=(\sqrt{n^0})^{-1}\!\cdot\!
{\hat Q}_{n^0}\!\cdot\!{\hat T}_-\!\cdot\!{\hat Q}_{n^0} \ , 
\qquad
{\hat T}_-=\sqrt{{\hat N}-n^0}\ {\hat c}\ , 
\nonumber\\
& &{\hat \gamma}_Q^*=(\sqrt{n^0})^{-1}\!
\cdot\!{\hat Q}_{n^0}\!\cdot\!{\hat T}_+\!\cdot\!{\hat Q}_{n^0} \ , \qquad
{\hat T}_+={\hat c}^*\sqrt{{\hat N}-n^0} \ , 
\nonumber\\
& &[{\hat \gamma}_Q , {\hat \gamma}_Q^* ]=2(n^0)^{-1}\!\cdot\!
{\hat Q}_{n^0}\!\cdot\!{\hat T}_0\!\cdot\!{\hat Q}_{n^0} \ , \quad
{\hat T}_0={\hat N}-n^0/2 \ .
\end{eqnarray}
The form of ${\hat T}_{\mp,0}$ is identical to the second Holstein-Primakoff 
boson representation of the $su(1,1)$-algebra.

\setcounter{equation}{0}

\section{SIMPLE EXAMPLE --- Damped and Amplified Oscillation ---}

\noindent
(1) The $su(1,1)$-algebraic model \cite{5} : 

The Hamiltonian for this model consists of 
\begin{eqnarray}\label{6-1}
& &{\hat H}={\hat K}_b-{\hat K}_a+{\hat V}_{ab} \ , 
\nonumber\\
& &{\hat K}_b=e\cdot({\hat b}^*{\hat b})+f\cdot({\hat b}^*{\hat b})^2 \ , 
\qquad
{\hat K}_a=e\cdot({\hat a}^*{\hat a})+f\cdot({\hat a}^*{\hat a})^2 \ , 
\nonumber\\
& &{\hat V}_{ab}=-ig\cdot({\hat b}^*{\hat a}^*-{\hat a}{\hat b}) \ . \quad
\end{eqnarray}
Here, $({\hat b}, {\hat b}^*)$ and $({\hat a} , {\hat a}^*)$ denote two 
kinds of bosons and $(e, f, g)$ are constants characterizing the Hamiltonian. 
The Hamiltonian (\ref{6-1}) can be expressed in the form 
\begin{equation}\label{6-2}
{\hat H}_{su(1,1)}^{(0)}=
2(e-f)\cdot({\hat T}-1/2) 
+4f\cdot({\hat T}-1/2)\cdot{\hat T}_0-ig\cdot({\hat T}_+-{\hat T}_-) \ .
\end{equation}
The set $({\hat T}_{\pm,0})$ obeys the $su(1,1)$-algebra and commutes 
with ${\hat T}$ : 
\begin{eqnarray}
& &{\hat T}_-={\hat a}{\hat b} \ , \quad
{\hat T}_+={\hat b}^*{\hat a}^* \ , \quad
{\hat T}_0=({\hat a}^*{\hat a}+{\hat b}{\hat b}^*)/2 \ , \qquad
\label{6-3}\\
& &{\hat T}=({\hat b}{\hat b}^*-{\hat a}^*{\hat a})/2 \ .
\label{6-4}
\end{eqnarray}
Since ${\hat T}$ commutes with ${\hat T}_{\pm, 0}$, ${\hat T}$ gives us a 
constant of motion. 
Then, in the space where the eigenvalue of ${\hat T}$ is equal to $T$, 
the Hamiltonian (\ref{6-2}) can be written as 
\begin{equation}\label{6-5}
{\hat H}_{su(1,1)}=
2(e-f)\cdot(T-1/2)
+4f\cdot(T-1/2)\cdot{\hat T}_0-ig\cdot({\hat T}_+-{\hat T}_-) \ .
\end{equation}

The Hamiltonian (\ref{6-5}) enables us to describe the damped and the 
amplified oscillation. The boson $({\hat a} , {\hat a}^*)$ plays a 
role of phase space doubling in the sense of the thermo field dynamics 
formalism presented by Umezawa et al.\cite{6} 
and it does not mean any physical object. 
Therefore, the Hamiltonian (\ref{6-5}) does not mean the total energy 
of the system and, then, 
it may be very difficult to draw a picture of motion induced by the 
Hamiltonian (\ref{6-5}). 
Of course, the results obtained in this model are quite interesting.

\noindent
(2) The $q$-deformed model : 

This model starts in the following Hamiltonian : 
\begin{eqnarray}\label{6-6}
& &{\hat H}={\hat K}_d+{\hat K}_c+{\hat V}_{cd} \ , 
\nonumber\\
& &{\hat K}_d=\omega\cdot({\hat d}^*{\hat d}) \ , \qquad
{\hat K}_c=\epsilon\cdot({\hat c}^*{\hat c})
=\epsilon\cdot{\hat c}^*\cdot[2/(1+[{\hat c} , {\hat c}^* ])]\cdot{\hat c} \ ,
\nonumber\\
& &{\hat V}_{cd}=-i\eta\cdot({\hat c}^*{\hat d}-{\hat d}^*{\hat c}) \ . 
\end{eqnarray}
Here, $({\hat d} , {\hat d}^*)$ and $({\hat c} , {\hat c}^*)$ denote 
two kinds of bosons and $(\omega, \epsilon, \eta)$ are constants 
characterizing the Hamiltonian.

Let Hamiltonian given in the relation (\ref{6-6}) deform by the 
function $f_0(n)=(\sqrt{1+n/n^0})^{-1}$ for 
the boson $({\hat c} , {\hat c}^*)$ : 
\begin{equation}\label{6-7}
{\hat H}_{{\rm def}}^{(0)}=
\omega\cdot({\hat d}^*{\hat d})+\epsilon\cdot({\hat c}^*{\hat c})
-i\eta\left({\hat c}^*\sqrt{1+{\hat N}/n^0}\cdot {\hat d}-{\hat d}^*\cdot
\sqrt{1+{\hat N}/n^0}\ {\hat c}\right) \ .
\end{equation}
For the Hamiltonian (\ref{6-7}), the following picture can be drawn : 
The external environment, for example, such as the heat bath, is 
described by the boson $({\hat d} , {\hat d}^*)$. 
The oscillation described by the boson $({\hat c} , {\hat c}^*)$ 
is damped and amplified due to the interaction with the external 
environment which is assumed to be extremely big system. 
Therefore, $({\hat d} , {\hat d}^*)$ has no fluctuation around 
the equilibrium value and, then, 
$({\hat d}, {\hat d}^*)$ can be replaced by the time-independent 
$c$-number $(\delta , \delta^*)$. 
It may be performed by calculating the expectation value of 
${\hat H}_{\rm def}^{(0)}$ for the boson coherent state for 
$({\hat d} , {\hat d}^*)$.

With the aid of the above picture, the following Hamiltonian 
is derived : 
\begin{eqnarray}\label{6-8}
& &{\hat H}_{\rm def}=
\omega\cdot|\delta|^2+\epsilon\cdot({\hat c}^*{\hat c})
-i\eta\cdot\sqrt{|\delta|^2/n^0}\cdot
[{\hat c}^* e^{-i\phi}\sqrt{n^0+{\hat N}}-\sqrt{n^0+{\hat N}}\ 
{\hat c}\ e^{i\phi}] \ , 
\nonumber\\
& &\qquad
\delta=|\delta|e^{-i\phi} \ , \qquad \delta^*=|\delta|e^{i\phi} \ .
\qquad\qquad\qquad\qquad\qquad
\end{eqnarray}
Further, the following relations are set up : 
\begin{eqnarray}
& &\zeta=\eta\cdot\sqrt{|\delta|^2/n^0}\ , 
\nonumber\\
& &\maru{c}={\hat c}\ e^{i\phi} \ , \qquad
\maru{c}^*={\hat c}^* e^{-i\phi}\ , \qquad
\maru{N}=\maru{c}^*\maru{c}\ (={\hat N}) \ , 
\label{6-9}\\
& &\maru{T}_-=\sqrt{\!n^0+\!\maru{N}}\ \maru{c} \ , \quad
\maru{T}_+=\maru{c}^*\!\!\sqrt{\!n^0+\!\maru{N}} \ , \quad
\maru{T}_0=\maru{N}\!+n^0/2\ .
\label{6-10}
\end{eqnarray}
Then, the Hamiltonian ${\hat H}_{\rm def}$ can be expressed as 
\begin{equation}\label{6-11}
\maru{H}_{\rm def}=(\omega-\epsilon/2)\cdot n^0
+\epsilon\cdot\maru{T}_0-i\zeta\cdot(\maru{T}_+-\maru{T}_-) \ .
\end{equation}
The relation (\ref{6-10}) shows us that $(\maru{T}_{\pm,0})$ 
is the Holstein-Primakoff boson representation of the $su(1,1)$-algebra and 
the formal structure of $\maru{H}_{\rm def}$ is completely the same as the 
${\hat H}_{su(1,1)}$ given in the relation (\ref{6-5}).

\setcounter{equation}{0}

\section{CONCLUDING REMARKS}

As for concluding remarks, we mention two points (A) and (B). 

\noindent
(A) The $q$-deformation of the $su(2)$- and the $su(1,1)$-algebra in two kinds 
of boson operators $({\hat a} , {\hat a}^*)$ and $({\hat b} , {\hat b}^*)$ : 

\noindent
In the above cases, the deformations are characterized by functions 
$f_0(x)$ and $g_0(x)$ in the form 
\begin{equation}\label{7-1}
[x]_f=x f_0(x-1)^{-2} \ , \qquad
[x]_g=x g_0(x-1)^{-2} \ .
\end{equation}
(1) The $su(2)$-algebra and its $q$-deformation : 
\begin{eqnarray}
& &{\hat S}_-^0={\hat b}^*{\hat a} \ , \quad
{\hat S}_+^0={\hat a}^*{\hat b}\ , \quad
{\hat S}_0=({\hat a}^*{\hat a}-{\hat b}^*{\hat b})/2 \ , 
\label{7-2a}\\
& &{\hat S}=({\hat b}^*{\hat b}+{\hat a}^*{\hat a})/2 \ , 
\label{7-2b}\\
& &{\hat S}_-=\sqrt{[{\hat S}+{\hat S}_0+1]_f[{\hat S}-{\hat S}_0]_g}
\left(\sqrt{{\hat S}+{\hat S}_0+1}\right)^{-1}{\hat S}_-^0\left(
\sqrt{{\hat S}-{\hat S}_0+1}\right)^{-1} \ , 
\nonumber\\
& &{\hat S}_+=
\left(\sqrt{{\hat S}-{\hat S}_0+1}\right)^{-1}{\hat S}_+^0
\left(\sqrt{{\hat S}+{\hat S}_0+1}\right)^{-1}
\sqrt{[{\hat S}+{\hat S}_0+1]_f[{\hat S}-{\hat S}_0]_g} \ , 
\nonumber\\
& &[2{\hat S}_0]{}_q=[{\hat S}+{\hat S}_0]_f[{\hat S}-{\hat S}_0+1]_g
-[{\hat S}+{\hat S}_0+1]_f[{\hat S}-{\hat S}_0]_g \ , 
\label{7-3}
\end{eqnarray}
(2) The $su(1,1)$-algebra and its $q$-deformation : 
\begin{eqnarray}
& &{\hat T}_-^0={\hat b}{\hat a} \ , \quad
{\hat T}_+^0={\hat a}^*{\hat b}^*\ , \quad
{\hat T}_0=({\hat a}^*{\hat a}+{\hat b}{\hat b}^*)/2 \ , 
\label{7-4a}\\
& &{\hat T}=({\hat b}{\hat b}^*-{\hat a}^*{\hat a})/2 \ , 
\label{7-4b}\\
& &{\hat T}_-=\sqrt{[{\hat T}_0-{\hat T}]_f[{\hat T}_0+{\hat T}-1]_g}
\left(\sqrt{({\hat T}_0+{\hat T})({\hat T}_0-{\hat T}+1)}\right)^{-1}
{\hat T}_-^0 \ , 
\nonumber\\
& &{\hat T}_+={\hat T}_+^0
\left(\sqrt{({\hat T}_0+{\hat T})({\hat T}_0-{\hat T}+1)}
\right)^{-1}
\sqrt{[{\hat T}_0-{\hat T}]_f[{\hat T}_0+{\hat T}-1]_g} \ , 
\nonumber\\
& &[2{\hat T}_0]{}_q=[{\hat T}_0-{\hat T}+1]_f[{\hat T}_0+{\hat T}]_g
-[{\hat T}_0-{\hat T}]_f[{\hat T}_0+{\hat T}-1]_g \ . 
\label{7-5}
\end{eqnarray}
(3) The most popular form : 
\begin{equation}\label{7-6}
f_0(n)=g_0(n)=
\sqrt{(n+1)(q-q^{-1})/(q^{n+1}-q^{-(n+1)})} \ . 
\end{equation}
For example, $[2{\hat S}_0]_q$ and $[2{\hat T}_0]_q$ are given in the form 
\begin{eqnarray}
& &[2{\hat S}_0]_q=(q^{2{\hat S}_0}-q^{-2{\hat S}_0})/(q-q^{-1}) \ , 
\label{7-7a}\\
& &
[2{\hat T}_0]_q=(q^{2{\hat T}_0}-q^{-2{\hat T}_0})/(q-q^{-1}) \ . 
\label{7-7b}
\end{eqnarray}
(B) The generalized Schwinger boson representation for the 
$su(M\!+1)$- and the 
$su(N,1)$-algebra and its $q$-deformation : 
%

\noindent
With the aid of Ref.\cite{7}, 
it may be possible to generalize the $q$-deformed algebra in two kinds 
of boson operators. Certainly, we have 
the $q$-deformation of the $su(2)$- and the $su(1,1)$-algebra 
in four kinds of boson operators \cite{8}. 
It is an example of the $q$-deformation of the generalized Schwinger 
boson representation. 
With the help of this form, for example, thermal effects on the 
pairing rotation, the paring vibration and the intrinsic structure may 
be described.

\acknowledgements

Two of the authors (Y. T. \& M. Y.) should acknowledge to Professor 
J. da Provid\^encia, co-author of this report and 
the chair of this workshop, for inviting them to Coimbra for several 
times in which the present work was discussed. Further, the author (M.Y.), 
who was given a chance to talk in this workshop, wishes thanks to 
Professor N. Kumar and Professor F. Bary Malik for their interest 
in this work.

\end{document}